\documentclass[preprintnumbers,prd,twocolumn,nofootinbib,superscriptaddress]{revtex4-1}

\usepackage{amsmath,amssymb,bm,bbm,amsfonts}
\usepackage{physics}
\usepackage{graphicx,graphics}
\usepackage[dvipsnames]{xcolor}
\usepackage[colorlinks=true,linkcolor=blue,citecolor=blue,urlcolor=blue]{hyperref}
\usepackage{dsfont}
\usepackage{comment}
\usepackage{hhline,multirow}
\usepackage{dcolumn}
\usepackage{url}
\usepackage[normalem]{ulem}
\usepackage{mathrsfs}
\usepackage{latexsym}
\usepackage{amscd}
\usepackage{slashed}
\usepackage{mathtools}

\usepackage{CJKutf8}

\newcommand{\vq}{\vec{q}}
\newcommand{\vs}{\vec{S}}
\newcommand{\vv}{\vec{v}}

\def\be{\begin{equation}}
\def\ee{\end{equation}}
\def\bea{\begin{eqnarray}}
\def\eea{\end{eqnarray}}

\begin{document}

\preprint{ADP-26-17/T1314}

\title{Direct detection constraints on EFT operators for inelastic dark matter}

\author{X.~G.~Wang}
\affiliation{CSSM and ARC Centre of Excellence for Dark Matter Particle Physics, Department of Physics, Adelaide University, Adelaide, SA 5005, Australia}

\author{A.~W.~Thomas}
\affiliation{CSSM and ARC Centre of Excellence for Dark Matter Particle Physics, Department of Physics, Adelaide University, Adelaide, SA 5005, Australia}

\begin{abstract}
We convert existing exclusion limits on the elastic spin-independent (SI) dark matter--proton cross sections into upper bounds on the event rate, subsequently applying them to constrain other effective operators for inelastic scattering of light dark matter in the GeV mass region. We focus on three representative scenarios including the isoscalar, proton-philic, and the proton-philic spin-dependent inelastic dark matter model, which are well-motivated in phenomenological studies of dark matter detection.
\end{abstract}

\date{\today}

\maketitle

\section{Introduction}
Unraveling the nature of dark matter is among the most critical unsolved problems in contemporary physics~\cite{Cirelli:2024ssz}. Weakly interacting massive particles (WIMPs)~\cite{Bertone:2004pz, Jungman:1995df, Goodman:1984dc}, which remain amongst the most compelling dark matter candidates, have attracted considerable effort using  various detection strategies. It is therefore essential to explore any avenue that might
constrain the nature of these particles.

Amongst strategies of dark matter searches, direct detection plays a crucial role, as it offers the most unambiguous way to establish the existence and properties of dark matter. While the DAMA/LIBRA collaboration~\cite{Bernabei:2013xsa, Bernabei:2018jrt} has claimed to observe a dark matter signal that is consistent with the expectations of annual modulation, the COSINE-100~\cite{COSINE-100:2021zqh, COSINE-100:2025kbw} and ANAIS~\cite{Amare:2019jul, Amare:2021yyu} collaborations disfavor the DAMA signal as originating from WIMP interactions. Moreover, other direct detection experiments in the low energy region of nuclear recoil have returned null results~\footnote{However, the LUX-ZEPLIN (LZ) experiment has recently reported a high-energy nuclear-recoil candidate at 
$E_{\rm nr} = 248 \pm 23_{\rm stat} \pm 23_{\rm sys}\ {\rm keV}$ with a maximum local significance of $3.4\ \sigma$ significance~\cite{LZ:2026axp}.}. 

It has proven quite useful to study the phenomenology of WIMPs in a model-independent fashion. In particular, non-relativistic effective field theory (NREFT)~\cite{Anand:2013yka, Fitzpatrick:2012ix} has been widely applied to describe the interactions between DM and the Standard Model (SM) particles. While the DAMA modulation amplitude could be well fitted by elastic scattering through the standard spin-independent (SI) or spin-dependent (SD) interactions~\cite{Savage:2008er, Baum:2018ekm}, numerous null experiments strongly exclude these regions of the SI~\cite{CRESST:2019jnq, CRESST:2019axx, DarkSide-50:2022qzh, XENON:2019rxp, PandaX-4T:2021bab, LZ:2024zvo} and SD~\cite{LUX:2017ree, PandaX-II:2016wea, PICO:2019vsc, LZ:2024zvo} parameter space.

Whether a dark matter model is consistent with existing constraints or anomalies strongly depends on the form of its interactions and kinematics. In order to satisfy the combined constraints from dark matter relic density, direct/indirect detection and collider searches, recent analyses suggested that the WIMP-nucleon interactions may be governed by effective operators beyond the standard SI and SD ones~\cite{Dent:2015zpa, DEramo:2016gos, Wang:2025cth}. Moreover, inelastic dark matter models~\cite{Tucker-Smith:2001myb, Tucker-Smith:2004mxa} have also been proposed, which could evade the direct detection limits, as incident dark matter particles may lack the kinetic energy required for the up-scattering process. Specific fine-tuning models, such as the proton-philic spin-dependent inelastic dark matter (pSIDM) model~\cite{Kang:2018qvz, Kang:2018zld}, have been proposed to provide a possible explanation of the DAMA modulation amplitude,  in compliance with the constraints from null experiments.

Experimental collaborations have expanded searches accordingly. The Xenon collaboration~\cite{XENON:2017fdd} has placed exclusion limits on isoscalar dimensionless couplings for all elastic scattering EFT operators. The PandaX-II collaboration~\cite{PandaX-II:2018woa} also set upper limits for couplings of relativistic interactions, which lead to  linear combinations of the effective operators in a non-relativistic reduction. 
The LZ collaboration~\cite{LZ:2023lvz} has set upper bounds on isoscalar and isovector couplings for both elastic and inelastic DM, although only heavy WIMPs ($m_{\chi} > 400\ {\rm GeV}$) have been considered in the inelastic case.

In this work, we convert the exclusion limits on the spin-independent DM--proton cross sections to upper bounds on the event rate. We then apply these bounds to constrain other effective operators for inelastic scattering of light dark matter in the GeV region. We investigate three representative scenarios including isoscalar, proton-philic and the pSIDM models. Our results for the isoscalar couplings are complementary to the LZ limits in the heavy mass region~\cite{LZ:2023lvz}.

We begin with a brief review of the event rate in direct detection within the non-relativistic effective field theory (NREFT) in Sec.~\ref{sec:NREFT}. The exclusion limits for each effective operator for inelastic DM are then presented in Sec.~\ref{sec:results}. Finally, we summarize our conclusions in Sec.~\ref{sec:conclusion}.

\section{Reconstruction of Event rate}
\label{sec:NREFT}

\subsection{Effective operators}

In the nonrelativistic limit, the interaction between WIMPs and nucleons can be written in terms of effective operators~\cite{Anand:2013yka, Fitzpatrick:2012ix}, 
\be
{\cal L} = \sum_i (c^p_i {\cal O}^p_i + c^n_i {\cal O}^n_i)\, ,
\ee
where
\bea
\label{eq:operators}
\mathcal{O}_1 &=& 1_{\chi} 1_{N}\, ,\nonumber\\ \mathcal{O}_2 &=& (\vv^{\perp})^2\, ,\nonumber\\ \mathcal{O}_3 &=& i \vs_N \cdot \left( \frac{\vq}{m_N} \times \vv^{\perp} \right)\, ,\nonumber\\
\mathcal{O}_4 &=& \vs_{\chi} \cdot \vs_{N}\, ,\nonumber\\ 
\mathcal{O}_5 &=& i \vs_{\chi} \cdot \left( \frac{\vq}{m_N} \times \vv^{\perp} \right)\, ,\nonumber\\ 
\mathcal{O}_6 &=& \left( \vs_{\chi} \cdot \frac{\vq}{m_N} \right) \left( \vs_N \cdot \frac{\vq}{m_N}\right)\, ,\nonumber\\
\mathcal{O}_7 &=& \vs_{N} \cdot \vv^{\perp}\, ,\nonumber\\
\mathcal{O}_8 &=& \vs_{\chi} \cdot \vv^{\perp}\, ,\nonumber\\ 
\mathcal{O}_9 &=& i \vs_{\chi} \cdot \left( \vs_N \times \frac{\vq}{m_N} \right)\, ,\nonumber\\
\mathcal{O}_{10} &=& i \vs_N \cdot \frac{\vq}{m_N}\, ,\nonumber\\ 
\mathcal{O}_{11} &=& i \vs_{\chi} \cdot \frac{\vq}{m_N}\, ,
\eea
with $\vv^{\perp} = \vv + \vq / 2 \mu_{\chi N}$ and $\mu_{\chi N} = m_{\chi} m_N / (m_{\chi} + m_N)$. 

The operator ${\cal O}_2$ is usually discarded as it cannot be obtained from the leading order nonrelativistic reduction of a manifestly relativistic interaction. ${\cal O}_1$ and ${\cal O}_4$ are the standard spin-independent (SI) and spin-dependent (SD) operators, respectively.

\subsection{Event rate of direct detection}

The differential DM--nucleus scattering cross section is~\cite{Fitzpatrick:2012ix}
%
\be
\label{eq:dsigma-dq2}
\frac{d \sigma_{\chi T}}{d q^2} = 
\frac{1}{4 \pi v^2} \cdot \sum_{ij} \sum_{N,N'=p,n} c^N_i c^{N'}_j F^{N,N'}_{ij}(q^2,v)\, ,
\ee
where $q^2=2 m_T E_R$ is the momentum transfer with $E_R$ the nuclear recoil energy. $F^{N,N'}_{ij}(q^2,v)$ are nuclear response functions, which can be decomposed in terms of the dark matter velocity,
\be
F^{N,N'}_{ij}(q^2,v) = F^{N,N'}_{ij,0}(q^2) + v^2 F^{N,N'}_{ij, 2}(q^2)\, .
\ee
The differential event rate (per unit time) in direct detection is given by
\be
\label{eq:dR-dER}
    \frac{dR}{d E_R} = \frac{\rho_{\chi}}{m_{\chi}} \int_{v_{\rm min}} v f(\vec{v},t) \cdot 2 m_T \frac{d\sigma_{\chi T}}{d q^2} d^3 v\ ,
\ee
where $\rho_{\chi} = 0.3\ {\rm GeV/cm^3}$~\cite{Baxter:2021pqo}, and $f(\vv,t)$ is the dark matter velocity distribution in the lab frame, which is time dependent because of the motion of the Earth.
The minimum velocity for a given recoil energy $E_R$ is
\be
\label{eq:vmin-el}
v_{\rm min} = \sqrt{\frac{m_T E_R}{2 \mu^2_{\chi T}}}\, ,
\ee
with $\mu_{\chi T} = m_{\chi} m_T/(m_{\chi} + m_T)$ the WIMP-nucleus reduced mass.

The integral over the dark matter velocity can be performed,
\bea
\label{eq:eta-h}
\eta(v_{\rm min},t) &=& \int_{v_{\rm min}}  \frac{f(\vv,t)}{v} d^3 v\, ,\nonumber\\
h(v_{\rm min},t) &=& \int_{v_{\rm min}} v f(\vv,t) d^3 v\, ,
\eea
leading to
\bea
\label{eq:dR-dER}
    \frac{dR}{d E_R} &=& \frac{\rho_{\chi} m_T}{2 \pi m_{\chi}} \sum_{ij}\sum_{N,N'} c^N_i c^{N'}_j \Big[ F^{N,N'}_{ij,0} \eta(v_{\rm min},t) \nonumber\\
  &&  + F^{N,N'}_{ij,2} h(v_{\rm min},t) \Big]\, .
\eea
The differential event rate averaged over a year can be written as
\bea
\label{eq:dRbar-dER}
    \frac{d \overline{R}}{d E_R} &=& \frac{\rho_{\chi} m_T}{2 \pi m_{\chi}} \sum_{ij}\sum_{N,N'} c^N_i c^{N'}_j \Big[ F^{N,N'}_{ij,0} \overline{\eta}(v_{\rm min}) \nonumber\\
  &&  + F^{N,N'}_{ij,2} \overline{h}(v_{\rm min}) \Big]\, ,
\eea
where 
\bea
\overline{\eta}(v_{\rm min}) &=& \frac{1}{T_0} \int_0^{T_0} \eta(v_{\rm min},t) dt\, ,\nonumber\\
\overline{h}(v_{\rm min}) &=& \frac{1}{T_0} \int_0^{T_0} h(v_{\rm min},t) dt\, ,
\eea
with $T_0 = 1$ year. 

Within the standard halo model (SHM), the velocity integrals in Eq.~(\ref{eq:eta-h}) can be calculated analytically~\cite{Gondolo:2002np, Catena:2015vpa}.

\subsection{Event rate with a SI interaction}

For velocity and momentum independent interactions, the cross section becomes~\cite{Freese:2012xd}
\be
    \frac{d\sigma_{\chi T}}{d q^2}(q^2,v) = \frac{\sigma_0}{4\mu^2_{\chi T} v^2} F^2(q) \Theta(q_{\rm max} - q)\, ,
\ee
where $\Theta$ is the Heaviside step function, and the zero-momentum cross section, $\sigma_0$, for a spin-independent (SI) interaction is
\be
    \sigma^{\rm SI}_0 = \frac{4 \mu^2_{\chi T}}{\pi} \Big[ Z c_1^p + (A - Z) c_1^n\Big]^2\, ,
\ee
with $Z$ and $A$ being the atomic number and atomic mass number, respectively. Assuming iso-spin symmetry, i.e., $c^p_1 = c^n_1$, 
\be
\sigma_0^{\rm SI} = \frac{\mu^2_{\chi T}}{\mu^2_{\chi N}} A^2 \cdot \sigma_p^{\rm SI}\, ,
\ee
where $\sigma_p^{\rm SI} = 4\mu^2_{\chi N} (c^p_1)^2/\pi$ is the WIMP--proton cross section that is often quoted in direct detection experiments.

The nuclear form factor is commonly approximated by the Helm form factor~\cite{Helm:1956zz, Lewin:1995rx},
\be
F(q) = 3 e^{-q^2 s^2/2} \frac{\sin(q r_n) - q r_n \cos(q r_n)}{(q r_n)^3}\, ,
\ee
where $s\approx 0.9\ {\rm fm}$ and $r^2_n = c^2 + \frac{7}{3} \pi^2 a^2 - 5 s^2$ is the effective nuclear radius, with $a\approx 0.52\ {\rm fm}$ and $c\approx (1.23 A^{1/3} - 0.60)\ {\rm fm}$.

The yearly averaged differential event rate can be written as
\be
\frac{d \overline{R}}{d E_R} = \frac{1}{2\mu^2_{\chi N}} \frac{\rho_{\chi}}{m_{\chi}} A^2 \cdot \sigma_p^{\rm SI} \cdot F^2(q) \cdot \overline{\eta}(v_{\rm min})\, .
\ee

The total event number per year per detector mass ($1/({\rm year} \cdot {\rm kg})$) can be obtained by summing over all possible recoil energies,
\bea
\label{eq:Rbar}
\overline{R} &=& \int_{E_{\rm th}}^{\infty} \epsilon(E_R) \frac{d\overline{R}}{d E_R} d E_R \nonumber\\
&=& \frac{\rho_{\chi} A^2 \sigma_p^{\rm SI}}{2 \mu^2_{\chi N} m_{\chi}} \int_{E_{\rm th}}^{\infty} \epsilon(E_R) F^2(q) \overline{\eta}(v_{\rm min}) d E_R\, ,
\eea
where $E_{\rm th}$ is the detector threshold and $\epsilon(E_R)$ is the energy efficiency.

The exclusion limits on $\sigma^{\rm SI}_p$ from direct detection experiments~\cite{XENON:2019rxp, LZ:2024zvo, PandaX-4T:2021bab} can be converted to upper bounds of the event rate according to Eq.~(\ref{eq:Rbar}).

\subsection{Inelastic DM}

In the case of endothermic inelastic scattering  with mass splitting $\delta_m = m_{\chi^*} - m_{\chi}$, the effective WIMP--nucleon operators in Eq.~(\ref{eq:operators}) should be modified with the variable $\vv^{\perp}$ being replaced by~\cite{Barello:2014uda}
\be
\vv^{\perp}_{\rm inel} = \vv + \frac{\vq}{2 \mu_{\chi N}} + \frac{\delta_m}{|\vq|^2} \vq = \vv^{\perp} + \frac{\delta_m}{|\vq|^2} \vq\, ,
\ee
which satisfies $\vv^{\perp}_{\rm inel} \cdot \vq = 0$. 

For WIMP--nucleus scattering, 
\be
\label{eq:v_perp_inel}
|\vv^{\perp}_{T,{\rm inel}}|^2 = |\vv_T|^2 - \frac{|\vq|^2}{4 \mu^2_{\chi T}} \left( 1 + \frac{2 \mu_{\chi T} \delta_m}{|\vq|^2} \right)^2\, ,
\ee
where $\vv_T = \vv_{\chi_1} - \vv_{T,in}$ is the WIMP--nucleus relative velocity.

The minimum velocity in Eq.~(\ref{eq:vmin-el}) is generalised to
\be
\label{eq:vmin-inel}
v_{\rm min} = \sqrt{\frac{m_T E_R}{2 \mu^2_{\chi T}}} + \frac{\delta_m}{\sqrt{2 m_T E_R}} \, .
\ee
The nuclear response functions should also be modified by the replacement,
\be
v^2 - \frac{q^2}{4 \mu^2_{\chi T}} \to v^2 - \frac{q^2}{4 \mu^2_{\chi T}} \left( 1 + \frac{2 \mu_{\chi T} \delta_m}{ q^2} \right)^2\, .
\ee
The up-scattering process requires that the kinetic energy of the incoming dark matter particle should be larger than the mass difference,
\be
\frac{1}{2}m_{\chi} v^2 \ge \delta_m\, ,
\ee
or equivalently, 
\be
\Delta_m = \delta_m/m_{\chi}\le v^2/2 \sim {\cal O}(10^{-6})\, .
\ee
As examples, we will take 
$\Delta_m = 1\times 10^{-7}$, $5\times 10^{-7}$, and $1 \times 10^{-6}$ in our numerical analysis.

\section{Results}
\label{sec:results}
In this work, we choose the most stringent constraints on $\sigma_p^{\rm SI}$ from the LZ collaboration~\cite{LZ:2024zvo}.  The energy efficiency $\epsilon(E_R)$ is taken from the orange curve in Fig.2 of Ref.~\cite{LZ:2024zvo}). The reconstructed event rate is derived by summing over all stable Xenon isotopes weighted by their natural abundances.

In our analysis, we deal with one effective operator at a time. The nuclear response functions are derived using the {\it Mathematica} script in Ref.~\cite{Anand:2013yka}. We consider three representative scenarios: (1) isoscalar: $c^p_i = c^n_i$; (2) proton-philic: $c^n_i = 0$; and (3) pSIDM model: with $c_4^n/c_4^p = - 0.028$, which maximally suppresses the cross section.

The proton-philic case is of particular interest in the dark photon models with either vector~\cite{Wang:2025clh} or axial-vector~\cite{Wang:2025cth} couplings, where the upper limits of the isoscalar couplings $(c^s_1)^2$ and $(c^s_8)^2$ from direct detection have been relaxed by a factor of $A^2/Z^2$, as a good approximation.

\subsection{Isoscalar}

\begin{figure*}[!h]
\includegraphics[width=14.5cm]{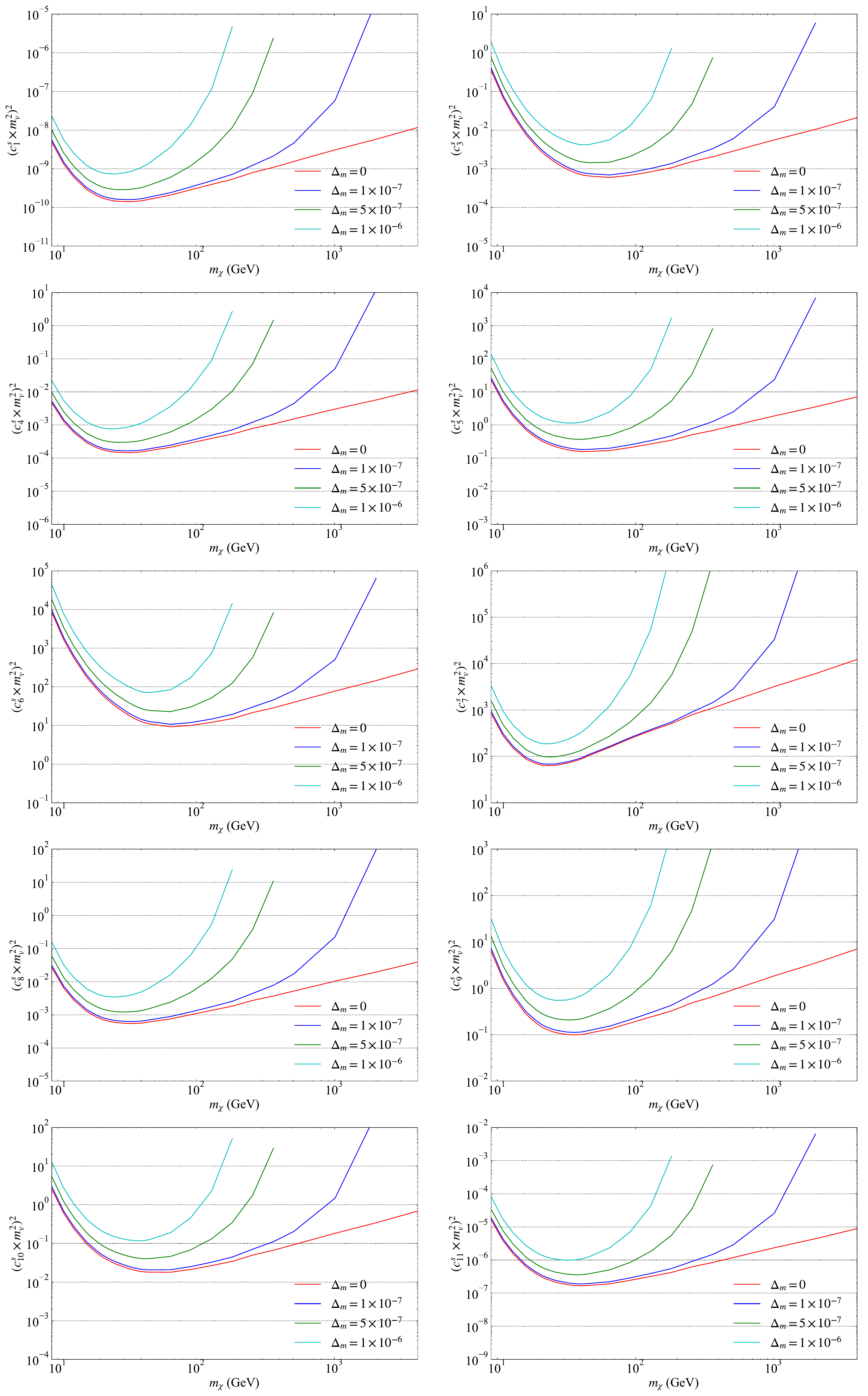}
\vspace*{-0.1cm}
\caption{The exclusion constraints of the dimensionless isoscalar WIMP-nucleon couplings for each effective operator,  with typical values of the mass splitting ratio, $\Delta_m = \delta_m/m_{\chi}$. These are derived by converting the LZ5T limits on the spin-independent (SI) WIMP-proton cross sections.}
\label{fig:cs_i_dimensionless_Deltam_abund}
\end{figure*}

In the case of $c^p_i = c^n_i$, we can define the isoscalar couplings $c_i^s = (c_i^p + c_i^n)/2$, and Eq.~(\ref{eq:dRbar-dER}) can be written as
\be
\label{eq:dRbar-dER-isoscalar}
    \frac{d \overline{R}}{d E_R} = \frac{\rho_{\chi} m_T}{2 \pi m_{\chi}} \sum_{ij} c^s_i c^s_j \Big[ F^{00}_{ij,0} \cdot \overline{\eta}(v_{\rm min})  + F^{00}_{ij,2} \cdot \overline{h}(v_{\rm min}) \Big]\, ,
\ee
where the isoscalar response functions are
\be
F^{00}_{ij,0(2)} = F^{p,p}_{ij,0(2)} + F^{p,n}_{ij,0(2)} + F^{n,p}_{ij,0(2)} + F^{n,n}_{ij,0(2)}\, .
\ee
The exclusion limits on the dimensionless isoscalar WIMP-nucleon couplings, $(c^s_i \times m_v^2)^2$ with $m_v = 246.2\ {\rm GeV}$, are shown in Fig.~\ref{fig:cs_i_dimensionless_Deltam_abund}, for both elastic ($\Delta_m = 0$) and inelastic scattering.

\subsection{Proton-philic}

\begin{figure*}[!h]
\includegraphics[width=14.5cm]{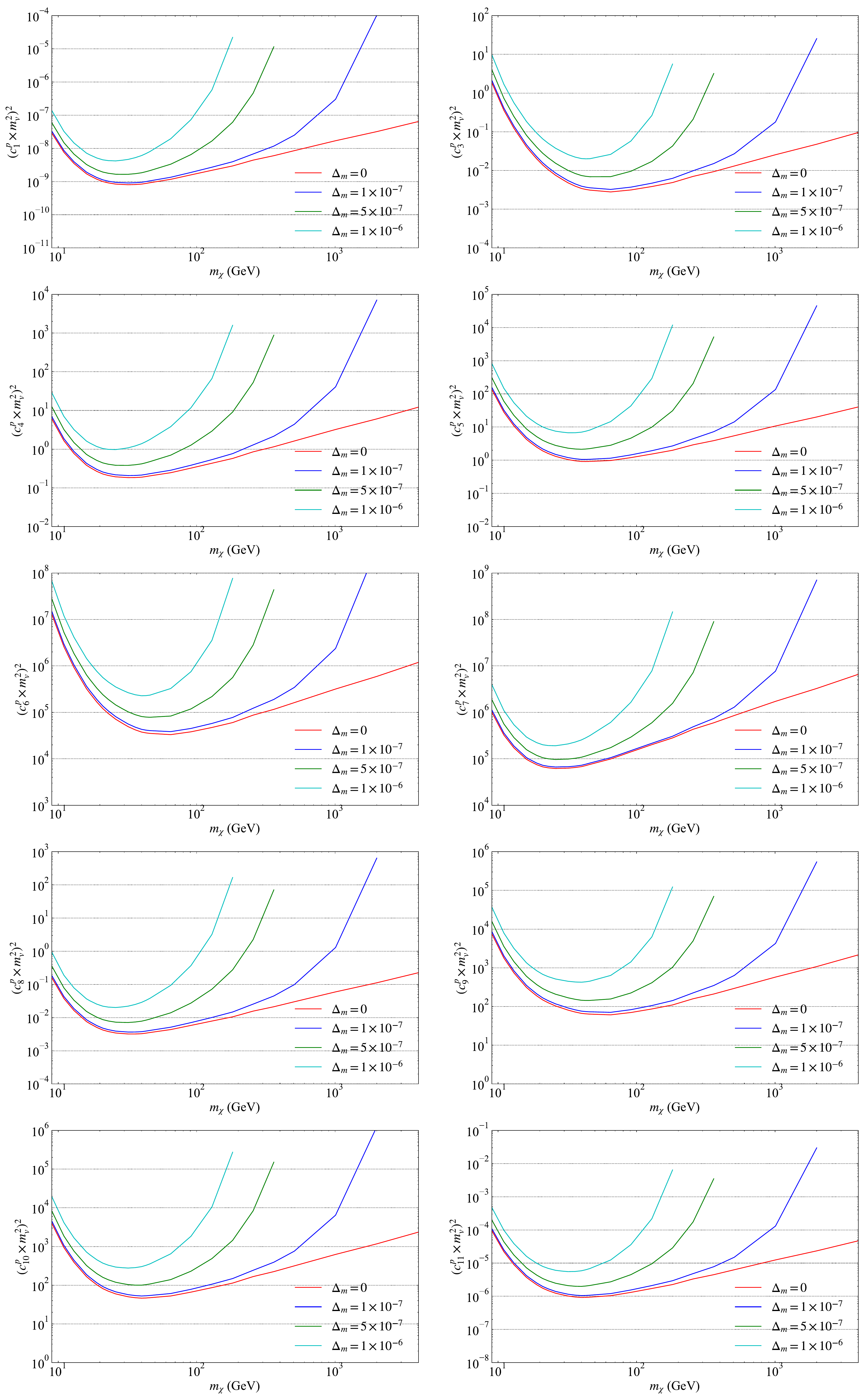}
\vspace*{-0.1cm}
\caption{The exclusion constraints of the dimensionless WIMP--proton couplings for each effective operators with typical values of the mass splitting percentage in the proton-philic scenario ($c_n=0$).}
\label{fig:cp_i_dimensionless_Deltam_abund_proton}
\end{figure*}

In the case where $c^n_i = 0$, Eq.~(\ref{eq:dRbar-dER}) can be written as
\be
\label{eq:dRbar-dER-proton}
    \frac{d \overline{R}}{d E_R} = \frac{\rho_{\chi} m_T}{2 \pi m_{\chi}} \sum_{ij} c^p_i c^{p}_j \Big[ F^{p,p}_{ij,0} \cdot \overline{\eta}(v_{\rm min})  + F^{p,p}_{ij,2} \cdot \overline{h}(v_{\rm min}) \Big]\, .
\ee
In Fig.~\ref{fig:cp_i_dimensionless_Deltam_abund_proton}, we show the resulting upper bounds on the DM--proton couplings. 

For effective operators which are independent of nucleon spins, such as ${\cal O}_1$ and ${\cal O}_8$, the upper limits are roughly consistent with a simple rescaling of the isoscalar couplings by a factor of $A^2/Z^2$.
For operators that depend on nucleon spins, (${\cal O}_{4,6,7,9,10}$), the upper limits on $(c^p_i)^2$ are significantly relaxed. This is because the corresponding nuclear response functions associated with protons are negligibly small, as the nuclear spins of $^{129}$Xe and $^{131}$Xe mainly arise from an unpaired neutron.

\subsection{pSIDM}

In the pSIDM model~\cite{Kang:2018zld}, the Xenon spin-dependent nuclear response functions are maximally suppressed by choosing fine-tuning parameters $c_4^n/c_4^p = - 0.028$,
\be
(c_4^p)^2 F_4^{(p,p)} + 2 c_4^p c_4^n F_4^{(p,n)} + (c_4^n)^2 F_4^{(n,n)} \ll (c_4^p)^2 F_4^{(p,p)}\, .
\ee
As a result, the exclusion limits on $c^p_4$ will be further relaxed compared with those with $c_4^n = 0$.

The best values of the parameters obtained by fitting the DAMA/LIBRA modulation amplitudes are~\cite{Kang:2018zld}
\bea
m_{\chi} &=& 12.1\ {\rm GeV}\, ,\ \ \delta_m = 18.3\ {\rm keV}\, ,\nonumber\\ 
\sigma_0^{\rm SD} &=& 7.95\times 10^{-35}\ {\rm cm}^2\, ,
\eea
where $\sigma_0^{\rm SD}$ is the DM--proton cross section which is related to the effective coupling $c_4^p$~\cite{Cirelli:2024ssz},
\be
\sigma_0^{\rm SD} = j_{\chi} (j_{\chi}+1) \frac{(c_4^p)^2 \mu^2_{\chi N}}{4\pi}\, ,
\ee
with $j_{\chi} =1/2$ the dark matter spin.

\begin{figure}[!t]
\includegraphics[width=\columnwidth]{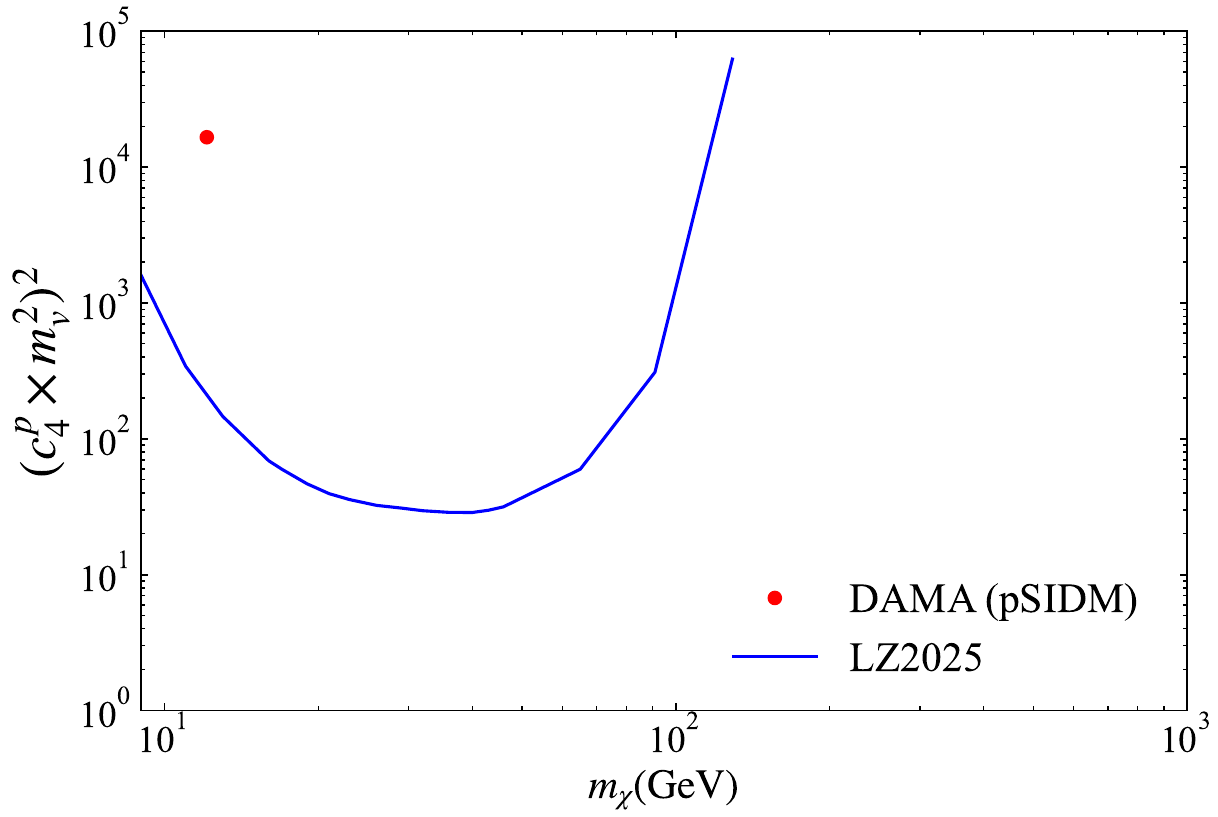}
\vspace*{-0.1cm}
\caption{The effective coupling to the proton in the pSIDM scenario, chosen to explain the DAMA modulation amplitude. The LZ2025 exclusion limits are derived using the parameters $c_n/c_p = - 0.028$ and $\Delta_m = 1.5\times 10^{-6}$.}
\label{fig:pSIDM}
\end{figure}

In Fig.~\ref{fig:pSIDM}, we show the resulting upper limits of $c_4^p$ with $c_n/c_p = - 0.028$ and $\Delta_m = \delta_m/m_{\chi} = 1.5\times 10^{-6}$. We can see that the DAMA signal is excluded by the LZ constraints even in the pSIDM scenario with highly fine-tuned parameters.

\section{Conclusions}
\label{sec:conclusion}
We have converted the exclusion limits on the spin-independent DM--nucleon cross sections obtained from direct detection to upper bounds of the event rate. These are then applied to constrain other effective interactions focusing on inelastic scattering of light dark matter in the GeV region. We present the exclusion limits of the couplings in three cases including the isoscalar, proton-philic, and the proton-philic spin-dependent inelastic dark matter scenarios. 

As we see from a comparison of Figs.~\ref{fig:cs_i_dimensionless_Deltam_abund} and \ref{fig:cp_i_dimensionless_Deltam_abund_proton}, the switch from isoscalar couplings to the proto-philic case relaxes the bounds on some couplings, notably $c_{4,6,7,9,10}$, by orders of magnitude. In the case of proto-phillic spin-dependent inelastic dark matter, fine tuned to maximally suppress the scattering cross section, the upper bound on $c_4$ is even higher. This is illustrated in Fig.~\ref{fig:pSIDM}. However, even in this extreme case the predictions for NaI are incompatible with results of the DAMA/LIBRA experiment. 

As a cautionary note, we observe that the constraints which we have obtained from direct detection could also be subject to some uncertainty from relativistic mean-field corrections~\cite{Wang:2020zgp}, changes in nuclear form factors~\cite{AbdelKhaleq:2023ipt} and dark matter velocity distributions~\cite{Kuhlen:2009vh}. Nevertheless, the results which have been obtained here may be profitably combined with constraints from indirect detection and collider searches and applied to further test various dark matter models.

\section*{Acknowledgments}
 We would like to thank Pengxuan Zhu and Giovani  Dalla Valle Garcia for helpful discussions. This work was supported by the University of Adelaide and by the Australian Research Council through the Australian Research Council Centre of Excellence for Dark Matter Particle Physics (CE200100008).


\bibliography{bibliography}

\end{document}